\documentstyle[aas2pp4]{article} 
\def\varphi{{\phi}}
\begin{document}

\title{The Interaction of New and Old Magnetic Fluxes at the Beginning of Solar Cycle 23}
\author{E. E. Benevolenskaya$^{1,2}$, J. T. Hoeksema$^{2}$,
A. G. Kosovichev$^{2}$, and P. H. Scherrer$^{2}$ }
\affil{$^{1}$Pulkovo Astronomical Observatory, St. Petersburg, 196140,
Russia}
 \affil{$^{2}$W.W.Hansen Experimental Physics Laboratory, Stanford University,
\\Stanford, CA 94305-4085}

\begin{abstract}
The 11-year cycle of solar activity follows Hale's law by
reversing the
magnetic polarity of leading and following sunspots in bipolar
regions during the minima of activity. In the 1996-97 solar minimum,
most solar activity emerged in narrow longitudinal zones - `active
longitudes' but over a range in latitude. Investigating the distribution
of solar magnetic flux,
we have found that the Hale sunspot polarity reversal first occurred
in these active zones. We have estimated the rotation rates of the
magnetic flux in the active zones before and after the polarity reversal.
Comparing these rotation rates with the internal rotation inferred
by helioseismology, we suggest that both `old' and `new' magnetic
fluxes were probably generated in a low-latitude zone near the base of
the solar convection zone.
The  reversal of active region polarity observed in certain longitudes
at the beginning of a new solar cycle suggests that the phenomenon
of active longitudes may give  fundamental information about the mechanism of the
solar cycle. The non-random distribution of old-cycle and new-cycle fluxes 
presents a challenge for dynamo theories, most of which
assume a uniform longitudinal distribution of solar magnetic fields.

\end{abstract}

\keywords{Sun: magnetic fields --- Sun: activity --- Sun: interior---Sun: sunspots---Sun: rotation}

\section{\bf Introduction}

The longitudinal and latitudinal distributions of
solar magnetic fields have been investigated by many authors
(e.g. Ambroz, 1973; Gaizaus\-kas, et al. 1983; Bumba, 1989; Bai, 1995).
Complexes of solar activity are  zones of field concentration 20$^\circ$--60$^\circ$ wide
that  during subsequent rotations tend to reappear at constant longitude
or drift slightly eastward or westward. These active zones  may persist
for 20--40 consecutive rotations and are called `Magnetic Active Longitudes'
(Bumba and Howard, 1969). According to Gaizauskas et al. (1983), complexes of
solar activity are maintained by new magnetic flux
contributed by active regions emerging within those complexes.
Moreover, each complex of solar activity rotates around the Sun at a steady rate.
The period is often close to the 27.28-day Carrington rotation period. 
A weaker background field pattern may rotate as a solid body at a different rate.

According to dynamo theory (e.g. Parker, 1993) the
source of the 11-year solar activity cycle is
located at the base of the convection zone, and the bipolar
sunspot groups are `$\Omega$-loops' of toroidal
magnetic field that erupt through the solar surface.
The toroidal $B_\varphi$ component is called positive when
its direction is westward, thus when the leading spot of a bipolar region
is positive, the $B_\varphi$ component is negative. 
During Cycle 22 the polarity of preceding sunspots was
positive in the southern hemisphere; the opposite was true 
in the northern
hemisphere where $B_\varphi > 0$. Cycle 23
began in September, 1996,
as determined from smoothed sunspot numbers.
Sunspot polarities in the new cycle
are reversed according to Hale's law
(Hale et al., 1919). Therefore, we
call the magnetic flux observed in Cycle 23 `old' if $B_\varphi > 0$ in
the northern hemisphere or $B_\varphi < 0$ in the southern hemisphere;
we call the magnetic flux `new' if the sign of  $B_\varphi$ is reversed.

The present study investigates the
latitudinal and longitudinal organization of the `new' and `old'
solar magnetic fields near solar minimum. We use synoptic maps derived from 
SOHO/MDI full-disk magnetograms
(Scherrer et al., 1995) obtained near  the
minimum of solar activity and during the beginning of the current
solar cycle. We analyzed data for 24 rotations beginning with
 Carrington Rotation 1911 (CR 1911),
 from June 1996 to April 1998. We compare the results
 with  data obtained during
the previous solar minimum, between May 1986 and February 1988,
at the Wilcox Solar Observatory (Hoeksema \& Scherrer, 1986).

\begin{figure}
\epsscale{1} \plotone{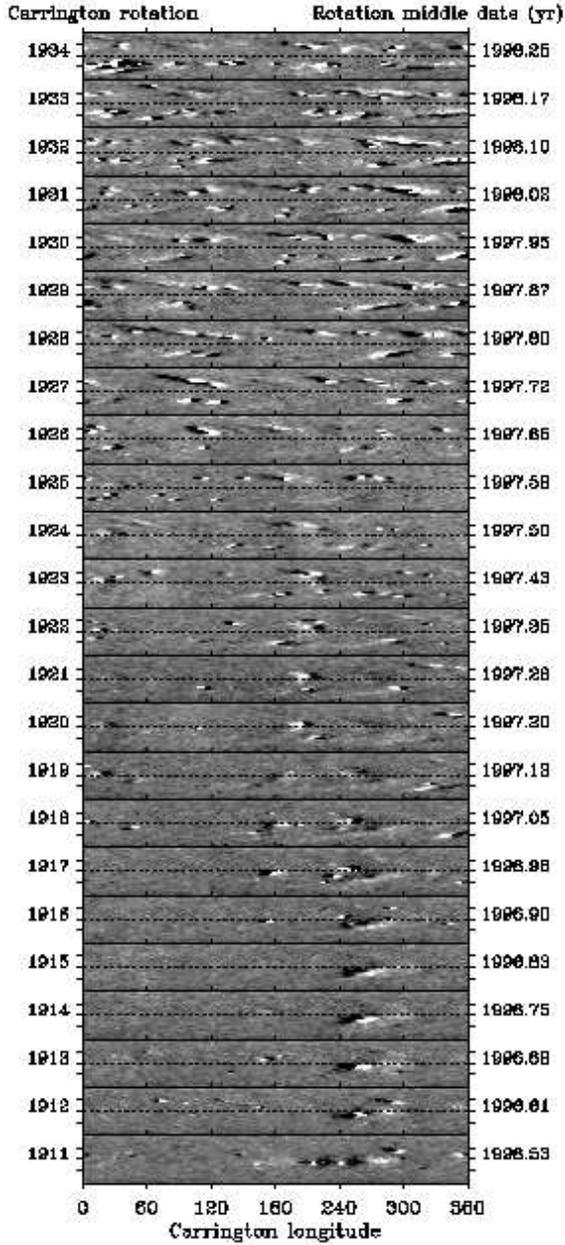}
\caption{Synoptic maps of the solar magnetic field for
Carrington Rotations 1911 to 1934 derived from the SOHO/MDI magnetograms
during the activity minimum between Cycles 22 and 23.
Values of the line-of-sight
component of the magnetic field  are
represented in light and dark colors for positive and negative polarities,
respectively. The grayscale shows magnetic field in the range from
$-10$ to 10 G.}
\end{figure}
\begin{figure}
\epsscale{1} \plotone{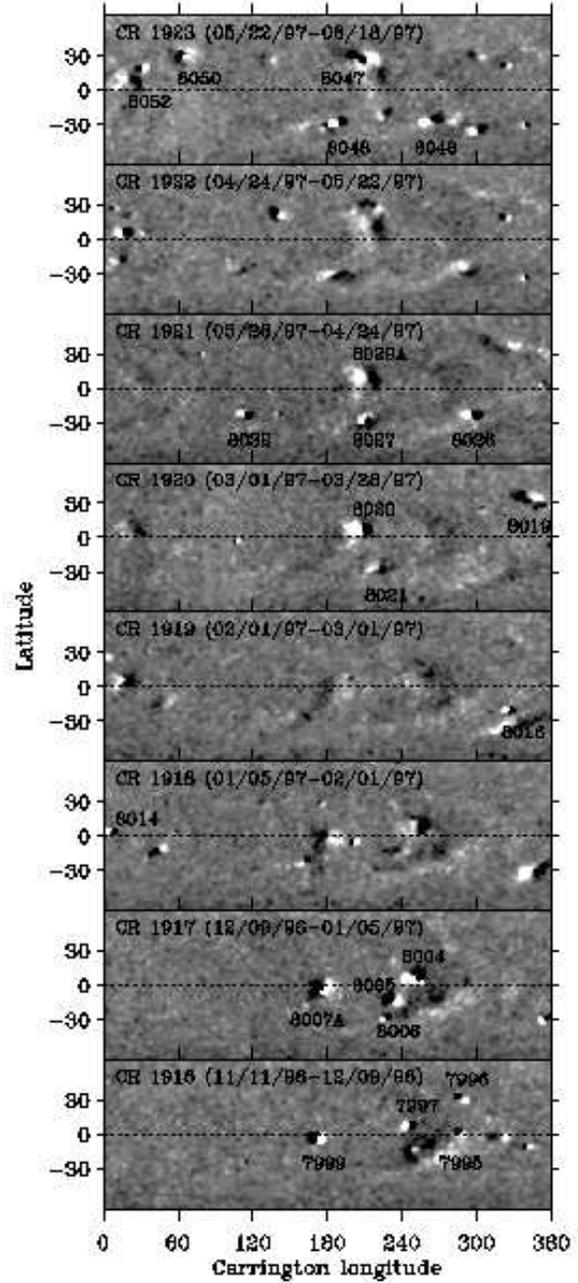}
\caption{More detailed synoptic magnetic maps for
Carrington Rotations 1916 to 1923. The plots indicate the NOAA
sunspot number for selected active regions. Bins are $1^o$ square
and extend to latitude $\pm 65^o$.}
\end{figure}

\section{\bf Results of the Analysis of Observations}
\begin{figure}
\epsscale{1} \plotone{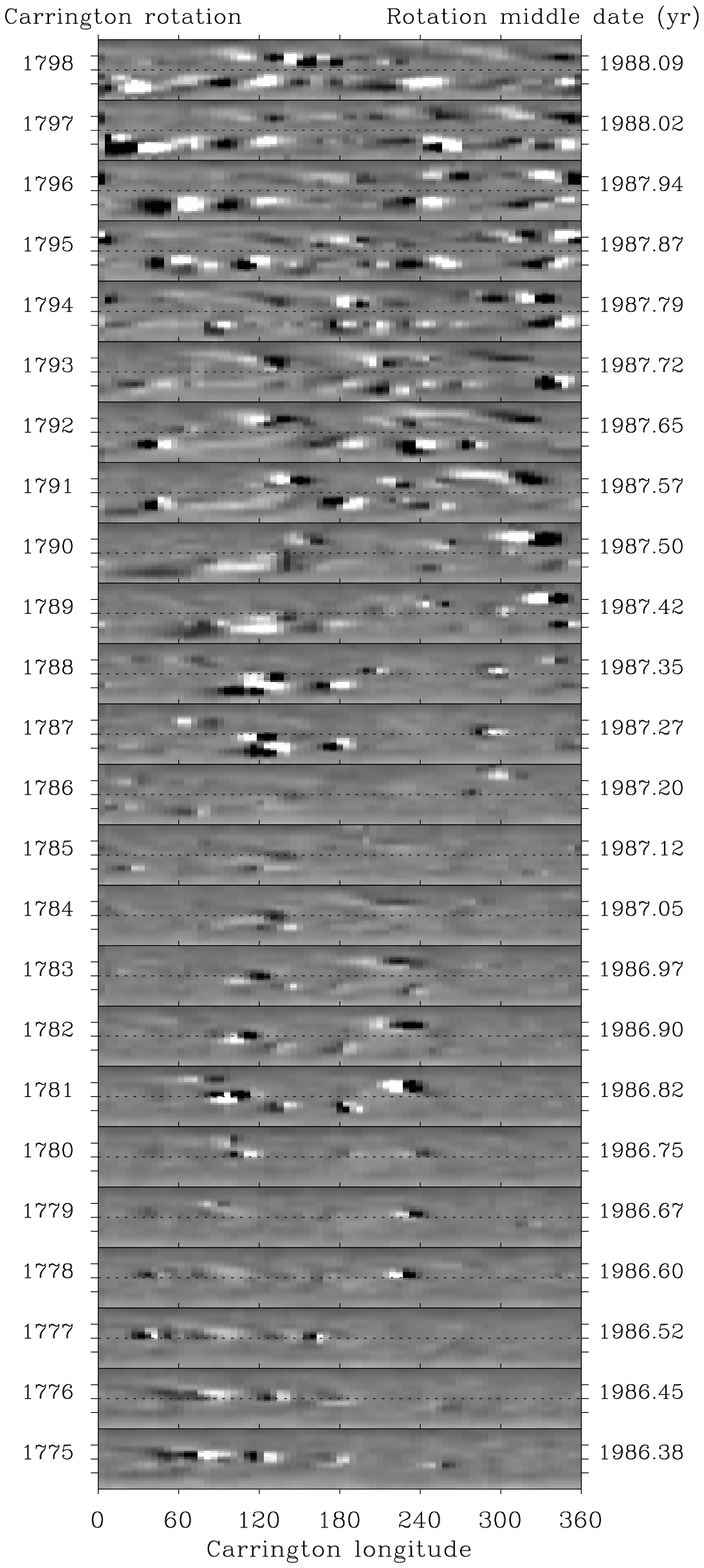}
\caption{Synoptic maps of the solar magnetic field for
Carrington Rotations 1775 to 1798 measured with the Wilcox Solar Observatory
magnetograph during the minimum between Cycle 21 and 22.}
\end{figure}

The SOHO/MDI full-disk magnetograms are $1024\times 1024$-pixel
images of the line-of-sight
component of magnetic field ($B_\Vert)$  measured in Gauss
on the solar surface. To obtain synoptic charts the original data were
binned down to $128\times 128$ pixels and then transformed into
the Carrington coordinate system.
The Carrington system is the  conventional longitudinal system on the Sun; 
it matches the  rotation rate of sunspots at a latitude of about
$16^\circ$ and has a mean synodic rotation period of about 27.2753 days.
The synoptic maps for CR 1911--1934 are shown
in Fig.~1.
The size of transformed elements on these maps
is 1$^\circ$ in both longitude and latitude. The maps cover from 0$^\circ$ to
360$^\circ$ in longitude and from -65$^\circ$ to 65$^\circ$ in
latitude.

During this interval the active regions are not distributed
randomly. They are clustered in complexes of activity, as in
earlier cycles (Bumba and Howard, 1969). Magnetic flux emerges
at particular fixed longitudes. During this period the `old' magnetic
flux of the previous solar cycle disappears at low latitudes
as `new' flux of the current cycle starts appearing at higher
latitudes ($\approx 30^\circ$).


The transition from old to new cycle flux is largely concentrated
in the interval CR 1916--1923,
which is shown in more detail in Fig.~2.
A few short-lived active regions of the new cycle
appeared much earlier.
For instance, during  CR 1911 a bipolar region of the new cycle ($B_\varphi
< 0$) in the northern hemisphere is visible at longitudes
280$^\circ$--300$^\circ$, but this region had
decayed by CR 1912. In CR 1912 a short-lived bipolar
region is visible in the northern hemisphere (30$^\circ$N) at
longitude 70$^\circ$.

We focus our investigation on zones living longer than  2 rotations.
One major zone occurred at longitudes
240$^\circ$--280$^\circ$ and lived over a year. 
This longitude was active from CR
1911 to CR 1917 (region NOAA 8006)
in the southern hemisphere and from CR 1916 to CR 1918 (NOAA 7997)
in the northern hemisphere. This active zone of old flux,
which was gradually decaying and migrating westward, reactivated in CR 1923,
when a new-cycle complex of solar activity (NOAA 8046)
 emerged in the southern hemisphere 
 at longitude $\approx
280^\circ$.

Another interesting strong
active zone developed at 160$^\circ$--200$^\circ$ and drifted slowly westward. 
This zone of
old-cycle flux first appeared in the southern hemisphere in CR 1916 (NOAA 7999)
and persisted
in CR 1917 (8007A); then in CR 1918 new cycle flux emerged in this zone at
latitude 20$^\circ$S.
During CR 1920 one of the last regions of the old
cycle appeared at longitudes 200$^\circ$--240$^\circ$ in the
northern hemisphere (NOAA 8020)
and decayed over the next two rotations; then, in CR 1923 a new cycle
region appeared in the same hemisphere but at
higher latitudes. In CR 1920 and 1921 both `new' (8021 and 8027)
and `old' (8020 and 8029A) fluxes existed at the same longitude
of $\simeq 210^\circ$, but in different hemispheres. We see a
similar coexistence of `new' (8006) and `old' (8005) regions
in the southern hemisphere in CR 1917.

In both active longitude zones the old-cycle magnetic flux was replaced by  new
cycle flux. The activity
in northern and southern hemispheres behaved differently, but
having the same active longitudes in the two hemispheres is quite common. 
A weaker active
zone  at 320$^\circ$--360$^\circ$ appeared in CR 1919, when
 activity complex of the new cycle (NOAA 8016) emerged at latitude 30$^\circ$S.
In CR 1920 the northern hemisphere became active instead of the southern
hemisphere at this longitudes (NOAA 8019).

After CR 1923 all significant bipolar regions were
of the new cycle and through CR 1934 the number of sunspots
substantially increased and covered latitudinal zones 10$^\circ$--40$^\circ$ wide in
both hemispheres.

During the previous solar minimum, between Cycles 21 and 22,
the interaction between the `old' and `new' magnetic fluxes was very similar:
the `old' flux was concentrated in two longitudinal zones, and
most of the initial `new' flux 
emerged in the same zones, at longitudes 230$^\circ$ and 120$^\circ$(Fig. 3).

\section{\bf Rotation of the Active Zones}

We have determined the sidereal rotation rate of  $4^\circ$-wide latitudinal zones
 near the equator (the `old' flux zone) and at $30^\circ$ 
(where the `new' flux appeared)
using the data for the current cycle.  
The rotation rate of magnetic flux was determined separately for
the northern and southern hemispheres.  To measure the rotation rate we
cross-correlated the magnetic fields measured for subsequent 
rotations in the latitudinal zones, then averaged the cross-correlation
functions separately for each zone 
for 10 rotations before CR 1922 and 10 rotations after CR 1923,
and, finally, determined the location of the maxima of the averaged 
cross-correlation functions
by fitting a Gaussian. We have also applied this procedure to the
individual longitudinal active zones using several different averaging periods 
and obtained similar results. The rotation rate of the equatorial zones in both
hemispheres ranged between 461.2 and 462.3 nHz, with a mean value
of 461.8 nHz 
and $1\sigma$-error estimate 0.5 nHz (corresponding sidereal period $25.06\pm 0.03$ days). 
This value is
close to the rotation rate of recurrent sunspots in this zone,
462.1 nHz (25.05 days), determined by Newton and Nunn (1951). For the $30^\circ$ zones,
the rotation rates in the northern and southern hemispheres were slightly
different: $446.6\pm 1.7$ nHz ($25.92\pm 0.10$ days) and $444.8\pm 1.6$ nHz 
($26.02\pm 0.10$ days) respectively.
Both these values are higher than the recurrent sunspot rotation 
rate of 440.6 nHz (26.27 days) at $30^\circ$ latitude.

Figure 4 compares these measurements with the internal
rotation rate inferred by helioseismology  (Schou et al., 1998). It appears that the
rotation rate of the old flux is close to the equatorial rotation
rate of plasma in the lower convection zone and in the upper part of the 
convection zone. However, the old flux 
rotates faster than the averaged rotation rate in the tachocline -
the region of the steep radial gradient of solar rotation at the bottom
of the convection zone, where  the magnetic flux is
probably generated (e.g. Golub et al., 1981; Parker, 1993). 
The rotation rate of the new flux
at $30^\circ$ latitude matches the internal rotation at the same
latitude only in a narrow zone of fast rotation 
in the upper part of the convection zone, approximately at 0.91--0.98 
solar radii; it is significantly lower than the rotation rate in the bulk of
convection zone including the tachocline.

\begin{figure}[t]
\epsscale{1} \plotone{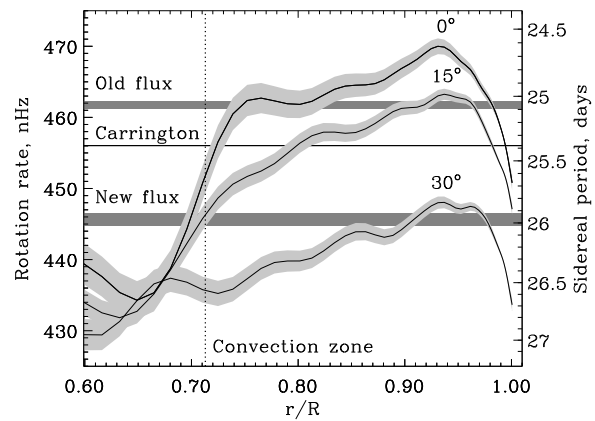}
\caption{Rotation rate inside
the Sun determined by helioseismology (Schou et al. 1998) as a
function of radius at three latitudes, $0^\circ$, $15^\circ$,
$30^\circ$  (solid curves with light gray areas
indicating $1\sigma$-error estimates).
The horizontal shaded areas
show the rotation rates of the old magnetic flux in the latitude range $1-5^\circ$,
and the new magnetic flux in the range $28-32^\circ$.
The horizontal solid line shows
the Carrington rotation rate, 456.03 nHz (sidereal period 25.38 days). The vertical dotted
line shows the lower boundary of the solar convection zone.
$R$ is the solar radius.
 }
\end{figure}
\section{\bf Discussion}

Solar activity was predominantly clustered in narrow longitudinal zones
during the recent transition from Solar Cycle 22 to 23, near the minimum of
solar activity.  Similar clustering occurred at the beginning of the previous
cycle.  New-cycle magnetic flux with reversed $B_\varphi$ polarity first
replaced old magnetic flux concentrations at the same longitude.  Only later
did the new-cycle flux emergence spread to other longitudes in both
hemispheres.  This observation suggests that the phenomenon of active
longitudes may provide fundamental information about the solar dynamo.
Current dynamo theories do not explain this phenomenon.  Ruzmaikin (1998)
recently suggested that longitudinal clustering of emerging magnetic flux may
be caused by large-scale non-axisymmetric modes of the $\alpha-\Omega$
dynamo.  It will be important to develop quantitative physical models of the
flux clustering.

The clustering of new and old cycle regions at particular longitudes suggests
that the primary dynamo regions are also localized in longitude.  The
subsurface rotation profiles (Fig. 4) suggest that this zone may also be
localized in latitude, viz. near the equator.

The new flux, which appears near $30^\circ$ latitude, rotates faster than the
mean convection zone rotation rate at that latitude.  It matches the
subsurface $30^\circ$ rotation only in a fairly narrow zone from 0.91-0.98 $R$.  
In principle flux originating at the bottom of the convection zone at
$30^\circ$ latitude could be accelerated to the observed rate in this rapidly
rotating sub-surface stream zone.  However, this is inconsistent with behavior
observed at low latitudes, where the rotation of the old flux at the equator
does not approach the rotation rate of the equatorial sub-surface stream at
the same depth.  

Perhaps the new flux comes from a different latitude and retains the rotation
rate of the deeper layer in which it originates.  Then one way to understand the
faster rotation of the new flux is if the flux generation occurs at a lower
latitude, where the rotation rate is higher, and the emerging loops are
transported to higher latitudes during their propagation through the
convection zone.

Suppose the new flux originates in a convectively stable layer just beneath
the convection zone, as indicated by most dynamo theories.  Then the 445 nHz
rotation rate suggests that the source layer is probably confined to low
latitudes, $15^\circ$ or even lower.  The near equatorial band also has the
strongest radial rotation rate gradient at the base of the convection zone,
something essential for the dynamo.  It will be important to study rotation
of the new and old fluxes in the active zones in more detail.

One possible mechanism for linking the locations of flux emergence associated
with the two cycles is if flux from different cycles resides at approximately
the same latitude and longitude but at different depths.  An irregularity at
a particular location associated with the dynamo of one cycle could perturb the
flux of the other cycle.  The result might be emergence of flux associated with
each cycle in both the northern and southern hemisphere at the same longitude.
Another possibility is influence of a non-uniform relict magnetic field
in the Sun's radiative interior (cf. Gough and McIntyre, 1998).

Our results make the difficult solar dynamo problem even more challenging,
because one must also explain why the old magnetic flux was replaced with new
flux of the opposite polarity in a narrow range of active longitudes during
the transition from one solar cycle to another.  We have also found that the
rotation rate of the new magnetic flux emerging at latitude $30^\circ$ was
higher than the rotation rate of the solar plasma in the bulk of the
convection zone.  This suggests that the new magnetic flux may be generated
in a narrow low-latitude zone, and maybe, even in the same near-equatorial
zone as the old flux.


\end {document}